\documentclass{aa}
\input epsf

\begin{document}
\sloppypar

   \thesaurus{06     
              (02.01.2;  
               02.09.1;  
	       08.02.3;  
	       08.06.3;  
	       08.14.1;  
               13.25.3)}  

   \title{Fourier power spectra at high frequencies: a way to
   distinguish a neutron star from a black hole}

   \author{R. Sunyaev \inst{1,2},  M. Revnivtsev \inst{2,1} }

   \offprints{revnivtsev@hea.iki.rssi.ru}

   \institute{Max-Planck-Institute f\"ur Astrophysik,
              Karl-Schwarzschild-Str. 1, D-85740 Garching bei M\"unchen,
              Germany,
	\and
  		Space Research Institute, Russian Academy of Sciences,
              Profsoyuznaya 84/32, 117810 Moscow, Russia       
             }
  \date{}

        \authorrunning{Sunyaev and Revnivtsev}
        \titlerunning{Fourier power spectra of neutron stars and black
        holes at high frequencies}
        
   \maketitle

   \begin{abstract}
	We analyzed the power density spectra of a sample of 9 neutron 
	star and 9 black hole binaries in the low/hard spectral state.
	In the power density spectra of accreting neutron
	stars with a weak magnetic field a significant power is
	contained at frequencies close to one kHz. 
	At the same time, most Galactic accreting black
	holes demonstrate a strong decline in the power spectra at
	the frequencies higher than 10--50 Hz. We propose to use this
	empirical fact as a method to distinguish the accreting
	neutron stars from black holes in X-ray transients. The X-ray
	transients that demonstrate significant noise in their X-ray flux at
	frequencies above $\sim500$ Hz should be considered neutron
	stars. We propose to explain the observed difference as a
	result of the existence of a radiation dominated spreading layer
	on the neutron star surface (Inogamov\& Sunyaev 1999).
	The possible very high frequency variabilities of this layer
	are discussed.

   \end{abstract}

%

\section{Introduction}
Among the Galactic X-ray sources black holes (BH) distinguish
themselves by the shape of their X-ray spectrum (see
e.g. \cite{tananbaum72}, \cite{shef_trump}, \cite{white_swank},
\cite{tanaka_shi}). In the low spectral state BHs emit a significant part of
their luminosity at energies of hundreds keV
(see e.g. \cite{two_src}), while neutron stars (NS) radiate much
smaller part of their total luminosity in this energy range (see
e.g. \cite{barret_review}). A soft component that is present in the 
spectra of BH binaries in the high(soft) state has a characteristic
temperature which can be significantly lower  
than that in the spectra of NS binaries with similar
luminosity. Besides, the BH
binaries in this 
soft spectral state demonstrate a hard power law tail (likely without
a high energy cutoff up to 500-600 keV, see e.g. Sunyaev et al. 1988,
1992; \cite{1655_osse}), whereas bright accreting 
NSs have never yet shown such spectra (e.g. \cite{tanaka_shi}).
 These spectral properties were frequently used as a
criterion to determine the nature of the compact object. We will not
discuss here the widely accepted methods of establishing neutron
star systems through the presence of pulsations or X-ray bursts. The
detection of coherent pulsations indicates 
the presence of the strong magnetic field and rotation of the NS. X-ray
bursts (type I) demonstrate that nuclear explosions occur in the
matter that was collected at the surface of the NS during the
accretion. However there is a significant number of sources, mostly 
transients, for which neither X-ray pulsations nor X-ray bursts have been
observed. Below we propose another method of determining the
nature of the compact object based on its power density spectrum (PDS) at
the high frequencies f$>$10--100 Hz.

\section{Observations and data analysis}
\begin{table}[htb]
\caption{The used observations of NS and BH binaries \label{log}}
\tabcolsep=0.1cm
\vspace{0.3cm}
\begin{tabular}{lccc} 
\hline 
Source&Proposal&Dates&Ref.\\
\hline
\hline
\multicolumn{4}{c}{Neutron Stars}\\
\hline 
GX~354-0$^*$&P10073&Feb. 22- Mar.1, 1996\\
/low state/\\
GX~354-0$^*$&P10073 &Feb.15, 1996&\\
/high state/\\
4U0614+091&P30056&Mar. 13-20, 1998&1\\
4U1608-522&P30062&Mar.31-Apr.2, 1998\\
/low state/\\
4U1608-522&P30062&Mar.24-27, 1998\\
/high state/\\
SAX~J1808.4--3658&P30411&Apr., 1998&2\\
1E1724--3045/Ter2$^*$&P10090&Nov.5-8, 1996&3\\
GS~1826--24&P30054&Feb.-June 1998\\
4U1705--44&P20073&Apr.1, 1997\\
SLX 1735--269&P20170&Feb.-&4\\
	      &P20089& -Oct. 1997\\
KS 1731--260&P30061&2-6 Oct. 1998& 5\\
Cyg X-2&P30418&2-6 July 1998\\
\hline
\multicolumn{4}{c}{Black Holes}\\
\hline
Cyg X-1$^*$&P30157&Dec.1997-Feb.1998\\
GX~339-4&P20183&1997&6\\
GS~1354--644&P20431&Nov.19-22&7\\
GRS~1915+105&P20402&Feb.9, 1997&8\\
GRO~J1655--40&P20402&Aug.14, 1997&9\\
4U1630-47&P30172&May-Jun.1998&10\\
XTE J1748--288&&July 1998&11\\
GRS~1758-258&P10231&1996-&12\\
            &P10232&-1998\\
            &P20166&\\
            &P30149&\\
1E1740.7-2942&P10231&1996-&12\\
            &P10232&-1998\\
            &P20166&\\
            &P30149&\\
\hline
\hline
\end{tabular}
\begin{list}{}{}
\item[$^*$] -- the data from these observations were also used for
analysis of the radiation spectra

\item[] -- {\footnotesize [1] --\cite{ford0614}, [2] -- \cite{wvdk1808}, [3] --
\cite{olive_ter}, [4] --  \cite{wvdk1735}, [5] -- \cite{muno1731}, [6]
-- \cite{nowak_339}, [7] -- \cite{mikej_1354}, [8] --
\cite{tsp_1915_1}, [9] --  \cite{remillard_1655}, [10] --
\cite{tomsick1630}, [11] -- \cite{mikej1748}, [12] -- \cite{smith_1eand1758}}
\end{list}
\end{table}

\begin{figure*}
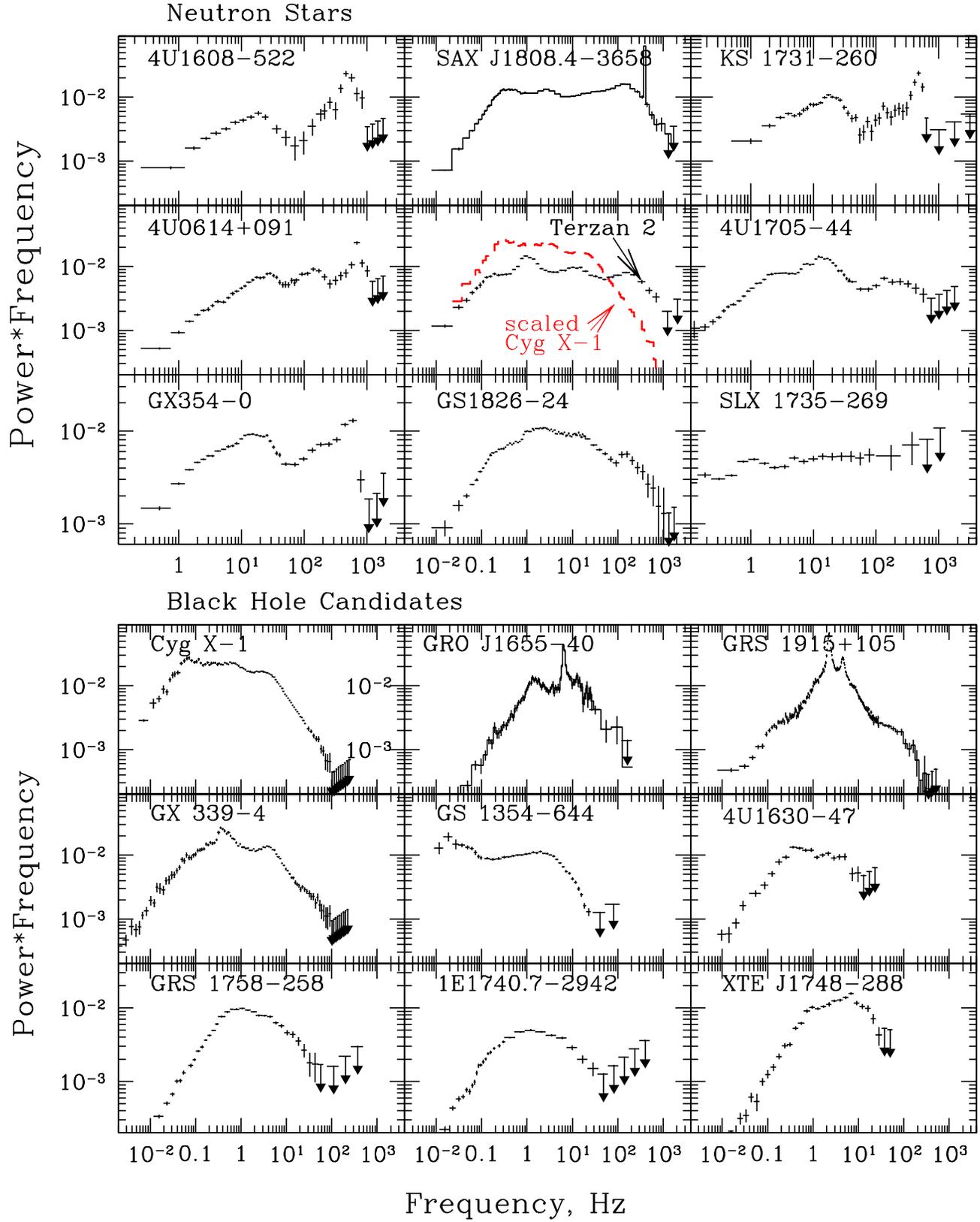

\vspace{-0.5cm}
\vbox{
\hspace{0.5cm}
\epsfxsize=18cm
\epsffile[40 150 600 520]{./h2161.f1a}
\vspace{-1.5cm}

\hspace{0.5cm}
\epsfxsize=18cm
\epsffile[40 150 600 520]{./h2161.f1b}
}
\caption{Broad band power spectra of X-ray binaries in the low
spectral state. Dashed line on the panel of Terzan2/1E1724-3045 shows
power spectrum of Cyg X-1 scaled according to their mass ratio
($f_{\rm Cyg X-1}\times7\rightarrow f_{\rm scaled~Cyg X-1}$). This
scaling is not enough to delete the discrepancy in the power
spectra of these sources.\label{powers}}. 
\end{figure*}

We used the publicly available data of the Proportional Counter Array
(PCA) aboard the Rossi X-ray Timing Explorer (RXTE) obtained in
1996--1998. We chose the observations of X-ray binaries in the
low/hard spectral state with $L_{\rm x}\sim10^{36}-10^{37}{\rm
erg/s}\sim0.01-0.1L_{\rm crit}$, where $L_{\rm crit}$ is the critical
Eddington luminosity, made in 1996-1998. We only used the data when all 5
Proportional Counter Units (PCUs) were operational.
The count rates for the sources vary
from a few hundreds counts/sec/PCA to several thousands
counts/sec/PCA for both NS and BH sources.
Our sample of NS and BH binaries is presented in Table \ref{log}.

We constructed the power density spectra of the X-ray sources using
$\sim$244 $\mu$sec ($2^{-12}$ sec) time resolution 
light curves divided into parts containing 8192 bins.
The obtained PDSs were normalized to the square of fractional variability
(e.g. \cite{miyamoto91}) and the ideal Poissonian noise component
($P_{\rm Poiss}(f)=2/R_{\rm tot}$) was subtracted from them. Then the obtained PDSs
were averaged. Fractional rms normalization is very useful for our
purposes because it has 
counting statistics noise component, with a  deviation from the ideal Poisson
level ($P_{\rm Poiss}(f)=2/R_{\rm tot}$) which is fairly independent of a source
count rate (see Eq. [\ref{eq:dead}]). To subtract the real (not ideal)
noise component due to the counting statistics from our PDSs 
we used the stationary Poissonian noise component level, modified by
the deadtime effects described in the papers of \cite{vihl94},
\cite{zhang_dt_95}, \cite{zhang_dt_96} and successfully tested in
e.g. \cite{morgan_1915_97}, \cite{nowak_cygx1} and \cite{jernigan}.

The explicit form of the used frequency dependent counting statistics
noise component, modified by deadtime effects appears as follows:

\begin{eqnarray}
P_{\rm dt}(f) &= &~~{{2}\over{R_{\rm tot}}}+ {{2}\xi\over{R_{\rm
     tot}}}\Bigg ( ~ \left [ ~-~ 2 {R_{\rm ph}} 
     \tau_{\rm d} ~ \left ( 1 ~-~ {{\tau_{\rm d}}\over{2 t_{\rm b}}} \right ) \right ]
     \nonumber \\
 & & -~  {{N-1}\over{N}} {R_{\rm ph}} \tau_{\rm d} ~
     \left ( {{\tau_{\rm d}}\over{t_{\rm b}}} \right ) ~ \cos \left (
     2 \pi t_{\rm b} f \right
     ) 
     \nonumber \\
 & &  + ~{ R_{\rm ph}}{ R}_{\rm vle} ~ 
      \left [ {{\sin \left( \pi \tau_{\rm vle} f \right )} \over {\pi f}}
      \right ]^2 \Bigg ) ~~,
\label{eq:dead}
\end{eqnarray}

\noindent
here $f$ -- the frequency; $R_{\rm tot}$ -- total (in our case -- for 5
PCUs) count rate of the  
source in the energy band of our interest; $R_{\rm ph}$ -- count rate of
the source in one PCU in energy band of our interest, in our case  
$R_{\rm tot}\approx 5 R_{\rm ph}$; $R_{\rm vle}$ -- count rate of Very
Large Events 
(VLE) in one PCU; $\tau_{\rm d}$ -- deadtime value for any detected event,
$\sim$9--10$\mu s$ (e.g. \cite{jahoda_dt1}, \cite{jahoda_dt2});
$\tau_{\rm vle}$ -- VLE preset 
window, $61\mu s$ (level 1) or $150 \mu s$ (level 2) (see
\cite{pca_hk}); $t_{\rm b}$ -- time binning in our light curves  
($\approx244 \mu s$); $N$ -- is the number of bins in the single light
curve segment ($N=8192$ in our case); $\xi$ -- is the
dimensionless parameter (very close to 1.0) that was introduced to roughly
take into account that the deadtime corrections for each PCU can be
not strictly identical and the model assumption of {\em stationary}
Poissonian process for the observed light curve is not exactly valid.
But in reality $\xi$ value is very close, within a few \%, to 1.0. The
closeness of $\xi$ to 1.0 can be the indication that all used model
parameters like $t_{\rm d}$ and $\tau_{vle}$ are close to the real ones.

 Following the conservative approach
adopted in \cite{jernigan} we fitted our PDSs at the highest frequencies
($f>600$ Hz for NSs and $f\ge50$ Hz for BHs) with the model for the white
noise component (Eq. [\ref{eq:dead}] without the first term, which was
already subtracted) and 
the model for the source itself, leaving the deadtime 
model normalization parameters $\xi$ and $R_{vle}$ free. We
used the simplest power law model for the intrinsic variability of the
sources. The deduced white noise parameters $\xi$ and $R_{vle}$ were
always consistent with the anticipated ones: $\xi\approx1.0$ (within a
few \%) and
$R_{vle}\sim100-180$ cnts/s/PCU. In the cases when the $R_{vle}$ (VLE
count rate) parameter was very poorly constrained (has large statistical
errors) we have it frozen at the values
determined from Standard\_1 mode data. This conservative approach helps us to
subtract most of the PDS components that are not real components of the
observed target. However, if the power spectrum of the observed source
has weak and very flat noise at frequencies of 500--2000 Hz it might be
treated as a noise component and will not be detected. In general, the
used approach for the deadtime modeling still has some uncertainties
such as the exact value of the 
deadtime for the photons $t_{\rm d}$, the stability of the VLE window
$\tau_{vle}$ and so on. Therefore we believe that our PDSs still
suffer from a small systematic uncertainty of the order of few$\times 10^{-7}$
(a few percent of $P_{N}(f)$) in units of squared rms.

Since we were mainly
interested in the high frequency 
continuum of the PDSs we paid special attention to all the known timing
features of the PCA instrument (see e.g. PCA Group Internet Homepage
http://lheawww.gsfc.nasa.gov/docs/xray/xte/pca). Apart from carefully
estimating the deadtime modification of the Poissonian level in our PDSs, we
ignored PCA channels 0--7 in all the analyzed data to avoid
background events variability (PCA Group Internet Homepage; E. Morgan,
private communication; see also \cite{jernigan}) that is especially
important for a weak NS sources like SLX~1735--269 and for black
holes because of their low intrinsic variability at high frequencies.  

To complete our power spectra with the low frequency part
($\sim10^{-2}-1$ Hz) we used Standard\_1 mode data (0.125 sec time
resolution). After the above procedure the power density 
spectra were renormalized to units of squared
relative rms {\em multiplied by frequency}. This helps to better visualize the
distribution of the power over the Fourier frequencies
(e.g. \cite{belloni_frms}). 

In Fig. \ref{powers} we present the power density spectra of 18
sources with subtracted
frequency dependent white noise level obtained with the described procedure.

We will present below a comparison of the radiation
spectra of Cyg X-1 and two NSs (Terzan2/1E1724-3045 and GX354-0). To
get them we used the same RXTE data as for the timing analysis (see Table
\ref{log}). The reduction of RXTE/PCA and RXTE/HEXTE data was done using
standard FTOOLS 4.2 tasks and the methods recommended by RXTE Guest Observer
Facility. 

\section{Results}

In Fig. \ref{powers} (upper part of plots) we present the obtained
power density spectra of nine X-ray 
bursters in the low/hard spectral state. For all of
these sources the nature of the compact object is known -- a NS with a
weak magnetic field. In the lower part of Fig.~\ref{powers}
we present the 
power density spectra for nine BH candidates. For two of them
(namely Cyg X-1, e.g. \cite{gies_bolton} and GRO J1655-40,
e.g. \cite{shahbaz_1655mass}) we 
have the mass determinations which indicate that these sources
can not be NSs. The other sources are believed to be black
hole binaries based on their spectral properties.

One can see 
the  striking similarity between the PDSs of the NS binaries and
BH candidates in the low spectral state up to $\sim$10 Hz,
which was already stated in the literature (see
e.g. \cite{wvdk_broadband}, \cite{psaltis99},
\cite{barret_review}). However, when we look at the high 
frequency part of the power spectra ($>$10--100 Hz) we notice that
even in the absence of kHz quasiperiodic oscillations (QPOs) there is
a dramatic difference between these two samples. The
PDSs of NS binaries have significant broad
noise components at the frequencies 500-1000 Hz, sometimes together
with well known kHz QPOs. The power density spectra of the BH 
binaries in the low spectral state (the typical state for Cyg
X-1) rapidly decrease at frequencies above 10 Hz and become by
one--two orders of magnitude lower than those of the NSs in the
same frequency band. The low frequency parts of the PDSs probably scale with
the compact object mass (see e.g. \cite{wvdk_broadband} or
Fig. \ref{powers}). However, it is likely that the high frequency part
of the PDSs of neutron stars demonstrates the additional noise component,
see Fig. \ref{powers}).

The observed empirical fact described above allows us to propose a new
method for establishing the nature of the compact object in bright 
transient X-ray sources which appear a few times a year in the X-ray
sky. 
We propose to classify sources whose PDS extends up to several hundred
Hz without a strong decline (in coordinates ``$frequency\times power$
vs. $frequency$'') as NSs with a weak magnetic field even if the kHz QPO
components are not detected in the PDS. 

\begin{figure}

\epsfxsize=8cm
\epsffile[25 170 550 700]{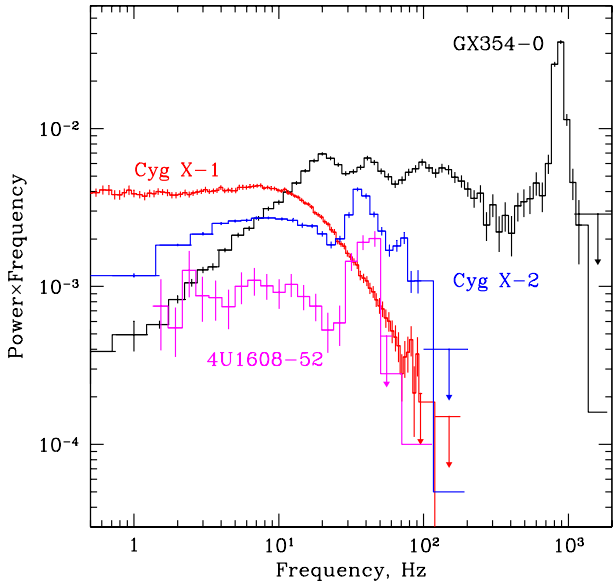}

\caption{Broad band power spectra of the black hole (Cyg X-1) and
NS binaries (GX354-0, Cyg X-2 and 4U1608--522) in the soft/high 
spectral state. The difference with Fig.~1 is clearly seen for at
least 4U1605--522 and Cyg X-2. GX~354-0 exhibits high frequency
variability even in the soft/high state.\label{softstate}}
\end{figure}

 We understand that the sample
of sources used in our analysis is not overwhelming. Moreover, the same X-ray
sources in the high spectral state usually demonstrate power spectra,
that may significantly differ (see Fig. \ref{softstate}) from each
other and from the low state PDS shown in Fig.\ref{powers}. In other words,
the NS systems in the high/soft spectral state could
demonstrate the absence of significant variability at high
frequencies. Therefore the absence of high frequency variability can
not be the reason to state that the source is a black hole candidate. 
However, we propose that any source that does demonstrate significant
variability (continuum or QPO) at frequencies close to 1 kHz should be
considered NS. Note that for four of the nine X-ray bursters
represented in Fig. \ref{powers}, kHz QPOs have not yet been observed.

\section{On the nature of the difference in PDS}  

\begin{figure}
\epsfxsize=8cm
\epsffile[25 150 570 690]{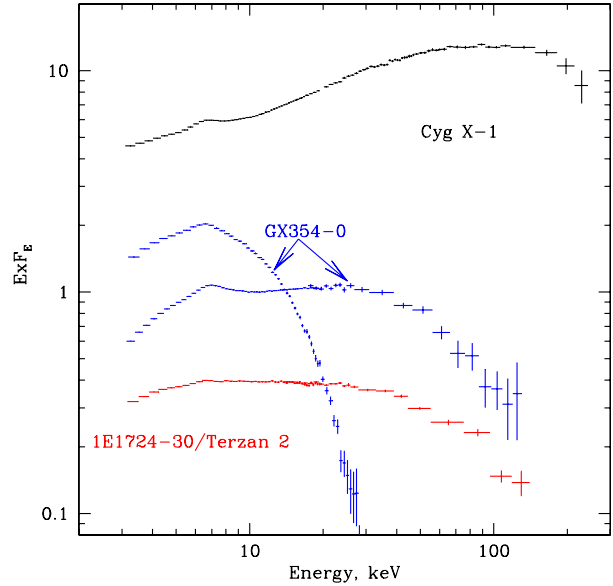}
\caption{The energy spectra of some of the analyzed X-ray binaries. The
source GX354-0 (4U1728-34) is shown in two spectral states - low/hard and
high/soft. Even in the hard state the spectra of NS are much softer
than the spectrum of Cyg X-1.\label{spectra}}
\end{figure}

The fate of accreting matter is very different in the case of
accretion onto a BH and NS with a small magnetic
field. In the case of an {\em accreting black hole} all
the observed radiation forms in an accretion disk, its corona  or in
an advection flow. In the case of a {\em neutron star} with a weak
magnetic field a significant part of the total 
gravitational energy should be released in the boundary layer
(e.g. \cite{ss88}, \cite{popham_bl}) or in a
layer of spreading matter {\em on the surface of the NS} (\cite{inogamov99}).
Approximately 2/3 of the total energy releases close to the star 
surface in the case of Schwarzschild geometry, see \cite{ss86}. Only about
$\sim$1/3 of the total 
energy is released in the extended accretion disk. If the neutron
star rotates rapidly these values change somewhat and the
accretion disk contribution increases (see e.g. \cite{ss99}). The surface
of the NS is an additional source of soft photons for
Comptonization. This might be the reason why the spectra of black
holes in the low/hard spectral state are generally harder than the spectra of
NSs (e.g. \cite{titar89}, \cite{two_src}, \cite{churazov_BHC}, see also Fig. \ref{spectra}).

\subsection{The instabilities originated in the accretion disk}

In the standard accretion disk, the region of the main energy release is a
resonator for acoustic and other instabilities with maximal
frequencies close to $\omega\sim c_{\rm s}/H\sim\omega_{\rm k}$, here $H$ -- the height of the accretion disk, $c_{\rm s}$ -- is the speed of sound,
$\omega_{\rm k}$ -- Keplerian frequency. In the case of
accretion on the BH with slow rotation the region of the main energy release
extends from $\sim5R_{\rm g}$ to $\sim27R_{\rm g}$, here $R_{\rm
g}=2GM/c^2$. For the heights $H<R$ and $M_{\rm BH}\sim10M_{\odot}$
there should be a maximal frequency $f\sim100$ Hz. Secular (Lightman \& Eardley
1974) and thermal (Shakura\& Sunyaev 1976) instabilities may develop
in the standard accretion disk but only in the region where  
the radiation pressure exceeds that of matter.
The characteristic time scales of these instabilities are 
considerably longer ($\tau\ga5/\alpha\omega_{\rm k}$, where
$\alpha\sim$ 0.01-1 is the viscosity parameter, see \cite{ss73}) than
the sound 
oscillation time scales mentioned above. One can anticipate that
the Velikhov-Chandrasekhar magnetorotational instability
(\cite{balbus}) could also result in the accretion disk variations on the 
time scales exceeding the orbital time.
An analysis of the disk turbulence
(e.g. \cite{nowak_wagoner95}) showed that a strong rollover above
$\sim$100 Hz should present in the power spectra of an accretion disk. The
phenomenological shot noise model (see e.g. \cite{terrel72},
\cite{lochner91}) and the model of coronal energy release variations
(e.g. \cite{poutanen_flares}) consider the similar absence
of strong variability at frequencies higher than 100 Hz. 
The advection dominated accretion flow (e.g. \cite{ichimaru},
\cite{narayan_yi}) are not subject to the thermal instability
mentioned above. In spite of the absence of detailed 
studies of different instabilities for the 
advection flows it seems quite reasonable 
to assume that a limit of the order of $H/c_{\rm s}$ may be valid in
this case as well. Rapid BH rotation can lead to increased
frequency values, but in this case the 
rotation parameter $a$ should be pretty close to the critical value 1.
If $a<0.5$ the influence of the black hole rotation is small.

All instabilities existing in the accretion disk modulate the flow of
matter onto the NS surface. Therefore, we could expect that
the majority of the types of variabilities we observe in accreting
BHs must manifest themselves in accreting NSs with
characteristic times proportional to the mass of the accreting object
(see e.g. \cite{ss76}, \cite{wvdk_broadband},
\cite{inogamov00}). However, in the BH systems the innermost regions 
of the accretion disk can generate the variability at its Keplerian
frequencies (close to $\sim$100--200 Hz), but {\em their contribution to
the observed PDS can be small because of the decrease of the energy
release close to the last stable orbit} (e.g. \cite{ss73}). In the NS
case, where the boundary layer can possess all accretion disk frequencies, this
variability could be more pronounced.

\begin{figure}

\epsfxsize=8cm
\epsffile[25 170 550 570]{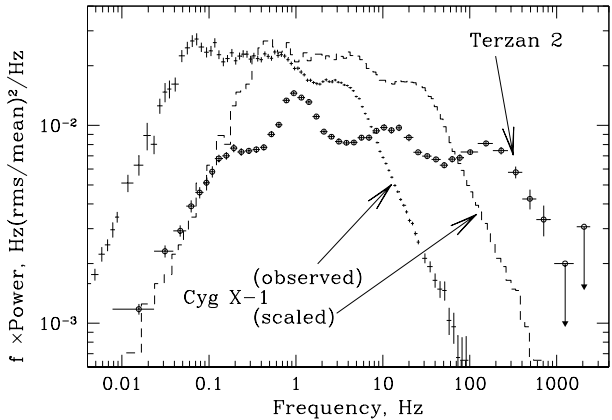}

\caption{The comparison of the power spectra of black hole (Cyg X-1)
and neutron star (Terzan 2). The dashed line shows the power spectrum
of Cyg X-1 scaled according to the mass ratio with Terzan 2
($f_{\rm Cyg X-1}\times7\rightarrow f_{\rm scaled~Cyg X-1}$). This
simple scaling is important but insufficient to 
explain fully the difference in the high frequency variability of
Terzan 2 and Cyg X-1. Note, that slopes of power density
spectra of Terzan 2 and Cyg X-1 in the high and low frequency limits
are similar, however the power spectrum of Terzan 2 is sufficiently
broader that that of Cyg X-1.  
\label{ter_and_cyg}}
\end{figure}

The simplest assumption is that the characteristics
frequencies in the power spectra of the sources scale as
$M^{-1}$. This scaling law is valid for e.g. Keplerian frequency in
the 
vicinity of the last stable orbit, thermal and secular instabilities
of the accretion disk in the region of main energy release,
Balbus--Hawley instability. However, this assumption is not enough to
account for the observed difference in the high frequency variability
of the NSs and 
the BHs. To demonstrate this we compared the PDS of the neutron star system 
Terzan2/1E1724--3045 ($M_1\sim1.4M{\odot}$) with $M_2/M_1$ times
scaled PDS of black hole system Cyg X-1 ($M_2\sim10M_{\odot}$)
(Fig. \ref{powers} and \ref{ter_and_cyg}). It is
obvious that such a scaling is important but insufficient to explain the high
frequency variability of Terzan2. It is most likely that there is an 
additional component coming from the neutron star surface or boundary
layer in its vicinity.

\subsection{The spreading layer and the highest possible frequencies in
the PDS of NSs}

The inner boundaries of accretion disks around BHs and neutron
stars lie at similar radii measured in the values of the gravitational
radii of the central object. However, we expect a small energy release
from the region with the radius smaller than the inner boundary of the
accretion disk around a BH. In the case of NS
accretion there is a solid surface and the matter should finally join
this surface. We know this because of the existence of X-ray
bursters. In accretion disks matter rotates with Keplerian
velocity. The velocity of the surface of the NS is
6--3 times lower (see e.g. \cite{stroh} or \cite{vdk_review} for
reviews). Therefore matter 
must decelerate and release its kinetic energy in the narrow spreading
layer on the  surface of the NS. In the case of very low
luminosities, less than 0.01 $L_{\rm crit}$ (where $L_{\rm crit}$ if the
critical Eddington luminosity) all deceleration occurs in the narrow
boundary layer between the innermost region of the accretion disk and
the surface of the NS. We could consider this layer an
atmosphere in which accreting matter is losing its kinetic energy and
radiating it away (see e.g. \cite{ss88}). At luminosities above 0.01
$L_{\rm crit}$ the radiation pressure does not permit such a simple
picture. Matter rotates around a NS and slowly 
spirals up the meridian toward higher latitudes. In this flow the
centrifugal force and the radiation pressure force balance gravitation
with very high precision. Effective gravity (the difference between
the gravitational force, centrifugal force and radiation pressure
force) becomes hundreds of times smaller than the gravitational
force. The turbulent friction with underlying denser layers rotating
with the surface velocity of the star leads to the deceleration of the
flow and energy release which is radiated away. This approach predicts
the existence of two broad bright belts equidistant from the star
equator. These belts are fed partially with the advection of thermal
energy from the regions closer to the equator
(\cite{inogamov99}). The luminosity of these bright belts exceeds the
luminosity of the accretion disk. 

The spreading layer may be considered a thin broad box. The flow along
the latitude has 
Keplerian velocity in the vicinity of the equator and slowly decelerates
to higher latitudes. This velocity is close to a hundred thousand
kilometers per second. The sound velocity $c_s$ within the radiation
dominated box is of the order of 20~000 km/sec. The highest sound
frequencies inside the 
box are of the order of $h/c_s$ and might reach even 40 kHz in the case of low
luminosities and 6-7 kHz in the case of luminosity close to the
$L_{\rm crit}$ (\cite{inogamov00}), here $h$ is the thickness of the
spreading layer. Simple consideration shows that we
could expect a lot of turbulence generated sound, resonant sound
waves and plasma instabilities with characteristic frequencies that do
not end at the Keplerian frequencies, but can reach several
kilohertz. 
The sound waves of much lower frequencies might also exist in the bright
belts: one can estimate the characteristic frequencies $f\sim c_s/(2\pi
R)\sim300$ Hz for the waves propagating 
along the latitude and $f\sim c_s/(R\Delta\theta)\sim$a few kHz (where
$\Delta\theta$ -- is the angular width of the bright belts) for the
waves propagating along the meridian within the belts.

As well as we expect the Keplerian frequency on the last stable orbit
($\sim200$Hz for $M_{\rm BH}\sim10M_{\odot}$) to be the
maximal possible frequency for the BH variability in X-rays, 40 kHz and
7 kHz mentioned above might be the the maximal frequencies for the
NSs PDS. In both cases (for BHs and NSs) the power spectra should drop
sufficiently towards these frequencies, but might be high below them.

In the spreading layer we have very strong shear in the flows along
the longitude. Under such circumstances the long living whirls like
those of thyphoons, Jupiter red spot or ``cats eyes'' due to
Kelvin-Helmholtz instability might regularly appear. The evolution of
the chaotic magnetic field structure inflowing into the spreading layer
through the neck of the accretion disk (region with $d\Omega/dr=0$)
may produce bright spots in the spreading layer.
The accretion disk screens from us half of the star, making it
completely unobservable. Part of the star surface is not visible
 due to eclipses by the star itself at any inclination angle except
$i$ close to 0. The 
rotation of the bright spots leads to their regular eclipses. The
duration of these eclipses is close to half of 
the period of the matter rotation in the spreading layer. The quasi
regular eclipses (the velocity of the bright spot can slowly change)
can also give rise to the additional variability (noise) of X-rays
from the NS at frequencies of hundreds of Hz.

There are many different mechanisms of instabilities in the narrow
(with heights from 400 m up to 1.5 km depending on the accretion rate
or luminosity), thin spreading layer.
Unfortunately, at the present level of sensitivity of RXTE we can not
see a significant variability of the sources at the frequencies above 
2--4 kHz. It is likely, that future missions with higher collecting area
of the detectors can shed light on the very high frequency
($f>5-10$ kHz) variability phenomena.  

\begin{acknowledgements}
Authors thank Nail Inogamov, Marat Gilfanov, Eugene Churazov and
Sergei Sazonov for helpful discussions. Also authors thank the
referee, Michiel van der Klis, whose comments helped us to improve the paper.
This research has made use of data obtained 
through the High Energy Astrophysics Science Archive Research Center
Online Service, provided by the NASA/Goddard Space Flight Center.
The work has been supported in part by RFBR grants RFFI 97-02-16264
and RFFI 00-15-96649.
\end{acknowledgements}

\end{document}